# Mobile and Residential INEA Wi-Fi Hotspot Network

Bartosz Musznicki, Karol Kowalik, Piotr Kołodziejski, and Eugeniusz Grzybek

INEA, Poznań, Poland
{bartosz.musznicki, karol.kowalik, piotr.kolodziejski, eugeniusz.grzybek}@inea.com.pl

*Abstract*—Since 2012 INEA has been developing and expanding the network of IEEE 802.11 compliant Wi-Fi hotspots (access points) located across the Greater Poland region. This network consists of 330 mobile (vehicular) access points carried by public buses and trams and over 20,000 fixed residential hotspots distributed throughout the homes of INEA customers to provide Internet access via the "community Wi-Fi" service. Therefore, this paper is aimed at sharing the insights gathered by INEA throughout 4 years of experience in providing hotspot-based Internet access. The emphasis is put on daily and hourly trends in order to evaluate user experience, to determine key patterns, and to investigate the influences such as public transportation trends, user location and mobility, as well as, radio frequency noise and interference.

## I. Introduction

There are currently about 47 million public Wi-Fi hotspots [1] around the globe. This form of Internet access is often provided by cafés, shops, airports, and railway stations to let the customers access the web, connect to the favourite social networking site, upload photos, read and send e-mails etc. Such an urban Internet access is also often used by tourists who do not use international data roaming packages, as well as, by teenagers when they have smartphones equipped with no or limited data packages provided by mobile network operators.

INEA is the largest regional fixed-access telecommunications operator in the Greater Poland, which provides advanced multimedia services to over 220,000 of homes, businesses, and institutions through different access mediums and technologies, i.e., Hybrid Fibre-Coaxial (HFC), Gigabit Passive Optical Network (GPON), point-to-point Carrier Ethernet optical fibres, IEEE 802.16e WiMAX [2], IEEE 802.11 Wi-Fi [3], as well as, twisted pair based xDSL and IEEE 802.3 Ethernet. INEA is also the major shareholder of the Greater Poland Broadband Network (*Polish:* Wielkopolska Sieć Szerokopasmowa [WSS]) that operates over 4,500 km of optical DWDM infrastructure with IP/MPLS architecture running on top of it to support Next Generation Access (NGA) services in 576 distribution nodes (points of presence), for the benefit of local communities, businesses, and administration.

Alongside the well-established services, INEA subscribers are offered free Wi-Fi hotspot access provided by over 20,000 fixed residential hotspots (with over 5,000 new hotspots planned in the in the upcoming months) and 330 mobile hotspots carried by public buses and trams [4]. There are hundreds of customers using the service each day, as depicted in Figure 1, which shows the changes in the number of users throughout an average day. Each data point corresponds to a mean value obtained in the period of one month (further discussed in Section II-D). As in every graph in the paper, confidence intervals (error bars) present the standard deviation of a data point. The graph for mobile network shows highest numbers of users in the morning and in the afternoon what corresponds to the busiest commuting hours. Stationary network, though, exhibits the highest usage in the evening what apparently relates to Internet activities performed at home.

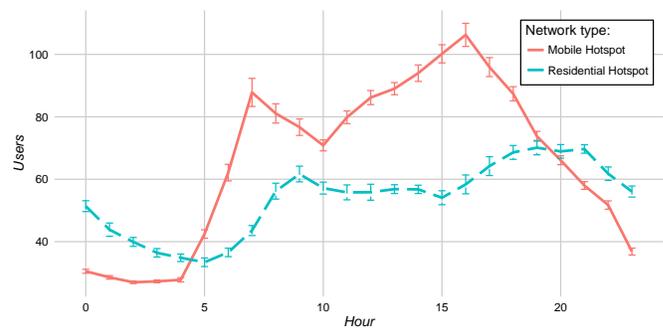

Figure 1. Number of INEA hotspot users throughout an average day

The rest of the paper is organized as follows. First, we give an overview of INEA mobile hotspot network, then we introduce the evaluation environment to move to experiment methodology and different signal quality related measurements. Later on, we go through similar steps with the residential hotspot network to present a range of user experience related tests and observations. Finally, we sum up with conclusions.

## II. Mobile hotspots

The INEA mobile hotspot network consists of 330 mobile access points (AP) mounted in public buses and trams, deployed in cooperation with municipal transportation operators of Poznań and Konin, i.e. Miejskie Przedsiębiorstwo Komunikacyjne (MPK) w Poznaniu [5] and Miejski Zakład Komunikacji (MZK) w Koninie [6], respectively. Each mobile AP device is a RouterBoard RB751U [7] enclosed within a protective enclosure and equipped with an external 2.5 dBi omni-directional antenna. Together with a cellular network based uplink (a 4G modem), those mobile access points provide wireless Internet access to the passengers and other users in the close proximity of the vehicle.

## A. User perspective

To use INEA mobile hotspot, one needs to select `INEA Hotspot` from the list of available networks, i.e. the list of Service Set Identifiers (SSID) detected by end user device. Before gaining Internet access, the user is redirected to a local captive portal (a landing page) with three options to access the service, as presented in Figure 2. Upon successful authentication the user is granted Internet access.

## B. Architecture and operation

Figure 2 provides an architectural overview of the mobile hotspot system. A centralized Remote Authentication Dial In User Service (RADIUS) database integrated with INEA's customer relationship management (CRM) system is used to perform authentication and authorisation. The RADIUS server also allows the service provider to obtain detailed information required by the data retention policy, such as, login and logout times, physical (MAC) addresses of the devices etc. Each access point is configured with a firewall and traffic control mechanisms responsible for policing connected devices. All mobile INEA hotspots share the same SSID, however they do not support network-coordinated roaming (handover), and hence, roaming has to be triggered by the user device. Should it take place, the user must re-authenticate but is not requested to manually select the access method again since session cookies are used to regain proper credentials and access.

The interaction with the system after the customer is presented with the captive portal depends on the selected option:
- free of charge access for INEA customers – the user is redirected to the site that performs authentication against INEA customers database,
- first 15 minutes free of charge access for visitors – a dedicated account is automatically created in RADIUS database to allow free 15 minutes during a 24 hour period,
- paid access for external users who require more than 15 minutes Internet access – the user is redirected to a payment website. After a successful payment, a dedicated RADIUS account is created with time limitations corresponding to the purchased package. The payment is realized by means of integration with *Przelewy24* system that provides wide range of payment methods for the Polish market, i.e. SMS payments, bank transfers etc [8].

Afterwards, when the authorisation phase has succeeded, the customer is allowed to access the Internet.

## C. Evaluation environment

After the initial tests, in order to perform a comparative study, not only 330 mobile access points have been used, but also a selected part of INEA's stationary 5 GHz Wi-Fi network – with similar number of concurrent customers (around 150), and with 10 stationary APs (base stations). Each stationary access point was a RouterBoard RB433 [9] enclosed within a protective enclosure and connected to a 16 dBi or 19 dBi external sector antenna.

## D. Experiment methodology

Each access point in the experiment has been queried using Simple Network Management Protocol (SNMP) every 15 minutes from 10 June 2016 till 10 July 2016. In this way, we have collected information about the noise floor perceived by each AP receiver, the number of connected clients, and the Received Signal Strength Indicator (RSSI) value for each connected client. The customers of these networks were not informed that the study was being performed, and hence, the collection of all the parameters which could be used to track customers (such as MAC addresses of customers Wi-Fi cards) was disabled. The 15 minutes interval was used in order to obtain the data frequently enough to get the full picture of changes in the performance throughout a full day, without affecting AP operations and customer experience in the process.

## E. Results

The comparison of mobile hotspots and stationary Wi-Fi access points is influenced by the patterns of people moving around the area and also by the number of active hotspots. Since most of the public buses are turned off for the night (except of night lines), while trams are still on, the number of active mobile hotspots drops in the night, as shown in Figure 3.

Around the noon, when there is a slightly lower demand for public transportation, and thus, some of the buses stay in bus depots, the decrease in the number of active hotspots can be observed. As a consequence, when analysing results in the following subsections, one should keep these patterns in mind.

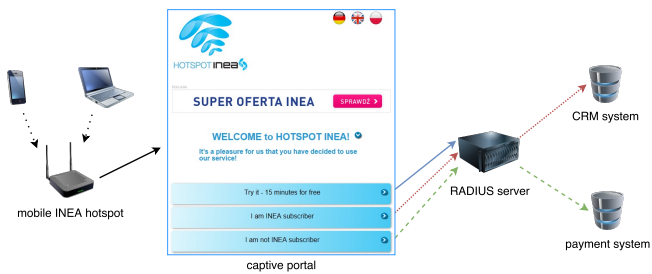

Figure 2. High-level architecture of INEA mobile hotspots authentication and authorisation system

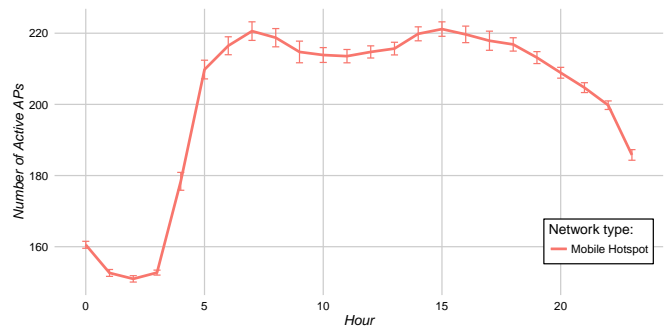

Figure 3. Number of active mobile INEA hotspots throughout a day

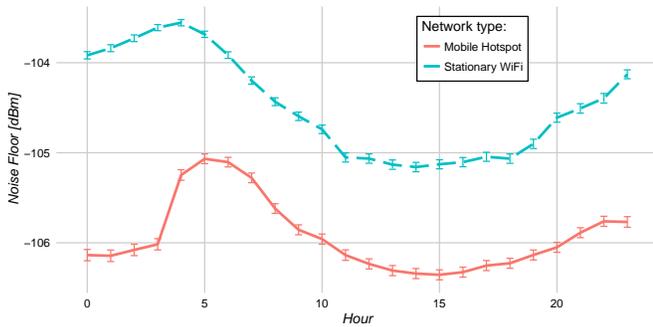

Figure 4. Average noise floor observed throughout an average day

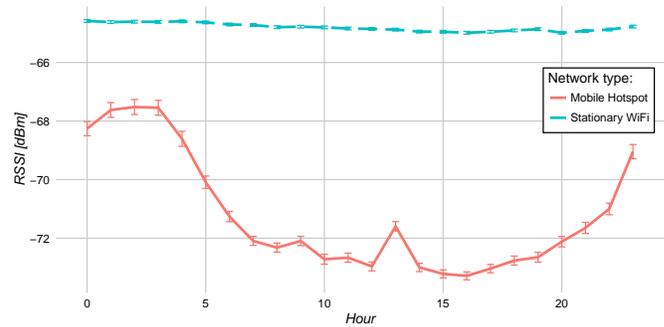

Figure 5. Average RSSI observed throughout an average day

*1) Noise floor:* The noise floor is the measure of background noise created by all the noise sources of the Radio Frequency (RF) environment. This ambient noise comes from variety of devices (microwave ovens, Bluetooth devices, wireless peripherals, ZigBee-based sensors, etc.), damaged connectors, radar equipment, as well as, atmospheric and thermal noise. Therefore, we investigate the noise floor values in order to verify if mobile hotspots exhibit similarities with stationary Wi-Fi networks.

In Figure 4 the average noise floor observed throughout an average day is presented, with the data points showing the mean value for a given hour. The data for both networks ware collected in different geographical locations and in different radio frequency bands (2.4 GHz for mobile hotspots and 5 GHz for stationary Wi-Fi) and on various radio channels spread over the set of available channels. This graph clearly illustrates the correlation between the noise floor for both mobile hotspot network and for stationary Wi-Fi network. Thus, it is valid to assume that the phenomenon is caused by sun activity or the processes taking place in the atmosphere. Obviously, the artificially caused influences exist as well, but due to averaging function performed over the whole month, they diminish on the presented diagram. Both curves exhibit similar characteristics, and yet, the one for stationary Wi-Fi network is around 2 dB lower than the mobile hotspot case. The difference is caused by the fact that stationary access points are operating on rooftops and are equipped with antennas of much higher gain, i.e. 16 dBi or 19 dBi, as compared to 2.5 dBi, and thus, they are exposed to noise approaching from wider areas then mobile hotspots. Although, the correlation between the two curves suggests that observed noise occurs due to natural activities and is not caused by mobility or changes in human-related patterns.

Our findings show the risks of drawing conclusions that noise floor solely depends on the human and device activities (spontaneous and intended radio transmissions). The results presented in Figure 4 imply that noise characteristic may be highly related to sun or atmospheric activities, while human-related activities contribute to the average level of noise.

*2) Received Signal Strength Indicator (RSSI):* This section studies the strength of the signal, expressed in the form RSSI from the client to the access point, and thus, all the reported measurements were gathered from the access points. The average RSSI observed throughout the day is presented in Figure 5. The client side hardware, i.e., Customer Premises Equipment (CPE), used in the stationary Wi-Fi scenario consists of an outdoor wireless receiver with a directional antenna. Moreover, most of the CPEs are located within the Line-of-Sight (LoS), so there are no obstacles which could cause significant variance in the observed RSSI values. Hence, not surprisingly, the average RSSI curve for stationary Wi-Fi is almost flat with a typical value around -65 dBm, which is considered a recommended value for the system.

The average RSSI for mobile hotspot network, on the contrary, changes significantly throughout the day. The high values during the night are most probably related to the trams which park at tram depots and provide Wi-Fi connectivity for the employees and the residents of the neighbouring areas. We assume that those users try to maximize signal strength staying nearby (in the optimum range), and thus, during the night the RSSI exhibits highest values. Similar situation occurs around 1 PM when some of the vehicles and drivers go back to bus depots, and again high RSSI values can be observed.

Figure 6 presents RSSI percentage histogram, which shows the percentage of samples collected for both types of networks for given RSSI value. The diagram is not cumulative thus the bars for mobile hotspot network cover (are presented in front of) the bars for stationary network. It demonstrates that stationary Wi-Fi is often actively used, and thus, we were able to collect much more samples than for the mobile hotspot network. Moreover, Figure 6 confirms that the stationary Wi-

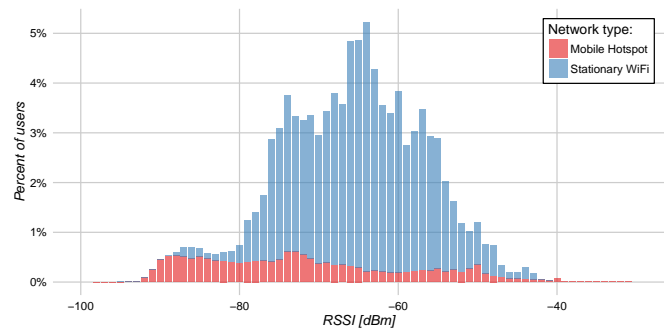

Figure 6. Distribution of mobile hotspots and stationary Wi-Fi RSSI

Fi network is designed and deployed with care. Therefore, most often observed RSSI values are around -65 dBm. On the other hand the mobile hotspot network attracts various types of users, which sometimes try to get and stay connected despite very low signal, as when on the fringes of the radio coverage range. Hence, the range of RSSI values for mobile network is much wider and typically around 10 dB below the stationary network. For this reason, the bitrates experienced by clients are often low, what as a consequence may occasionally lead to reduced overall capacity.

## III. Residential hotspots

INEA operates the network of over 20,000 residential hotspots that follow the idea commonly known as "community Wi-Fi" and provide Internet access to the members of the community. Hence, according to INEA's policy for the Data Over Cable Service Interface Specification (DOCSIS) access network, every Cisco EPC3925 cable modem (a CPE), equipped with an integrated IEEE 802.11n 2x2 MIMO 2.4 GHz Wi-Fi module [10] by default operates not only as a personal (home) Wi-Fi router, but also as a virtualized community access point. Each customer is allowed to opt out of this complimentary service, but as a result will no longer be allowed to access the distributed residential hotspot system [4]. This decision has no effect on other services the customer is subscribed to in a monthly fee model and still can enjoy home Internet access available in packages up to 250 Mbit/s.

### A. User perspective

To access the Internet via community Wi-Fi, INEA customers have to connect to the SSID called `INEA_HotSpot_WiFi` and login with the credentials they use to access INEA customer care portal. When the first authentication is successful, in most cases, depending on the configuration of end user device, the credentials input by the user are stored in the (mobile) device for further use. Therefore, the next time the customer turns on the Wi-Fi while being in the radio range of INEA residential hotspot system, the customer will be automatically logged in to the network without the need to repeat the process by hand. Moreover, Protected Extensible Authentication Protocol (PEAP) [11] based approach allows INEA to provide seamless authentication for customers that are moving between access points. For example, let us consider a customer who uses his or hers home Wi-Fi provided by INEA on a daily basis. When visiting a friend or a relative who is also a member of INEA Wi-Fi community (or lives nearby a community member), the user will be able to access the Internet through the local CPE. Therefore, regardless if the device is a smartphone, a tablet, or a notebook, it can connect to the network without the user being aware of the automated processes taking place in the background.

### B. Architecture and operation

Each INEA community Wi-Fi access point is in fact an additional service provided by a single 2.4 GHz IEEE 802.11n [12]

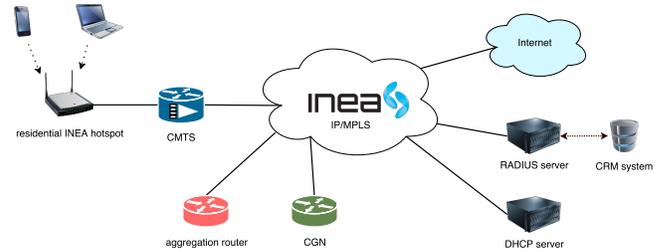

Figure 7. High-level architecture of residential INEA hotspot network

compliant radio module incorporated in the CPE. When booting, the wireless channel the AP will operate on is automatically selected from those that are the least used by other Wi-Fi devices in the radio range. Service differentiation and separation from the home Wi-Fi router is achieved by broadcasting an additional Basic Service Set Identifier (BSSID) with network name, i.e. SSID, set to `INEA_HotSpot_WiFi`. Then the logical interface associated with that BSSID is internally linked (within the CPE) to a dedicated DOCSIS 3.0 service flow [13], i.e. a MAC layer transport tunnel that logically separates hotspot traffic from home user's traffic in the last mile section between CPE and Cable Modem Termination System (CMTS). The data traffic related to the hotspot is limited to 2 Mbit/s (downlink) and 1 Mbit/s (uplink) according to the applied Quality of Service (QoS) policy. Therefore, the common Hybrid Fibre-Coaxial access medium in the last-mile of the DOCSIS-based network can be efficiently used to provide both high speed Internet access to the local (home) subscriber, as well as, to provide the community Wi-Fi for visitors.

Following the guidelines of RFC 6598 [14], a DHCP server assigns an IPv4 address which belongs to the 100.64.0.0/10 prefix for every end user device connected to the residential hotspot network and the data is further transmitted via CMTS in a Layer 2 Over Generic Routing Encapsulation (L2oGRE) tunnel to an aggregation router, as presented in Figure 7. Then the traffic undergoes the process of Network Address Translation (NAT) performed by INEA's Carrier-Grade NAT (CGN) system. To ensure uninterrupted and transparent services, the tunnel aggregation and CGN systems are built in a strict high availability architecture. Security measures applied to `INEA_HotSpot_WiFi` are separated from the mechanisms used by the home router Wi-Fi service. The authentication is based on IEEE 802.1X standard [15] and uses PEAP and RADIUS together with a centralized user database.

### C. Experiment environment

To tackle the assessment of user experience in the community Wi-Fi network, an approach opposite to the one applied in Section II has been used. Therefore, an empirical study involving end user devices has been performed to evaluate the actual usefulness and quality of the service. The methodology was twofold: first to investigate the experience of a smartphone user who moves through the test area, secondly to evaluate the service when a user keeps a fixed position inside an apartment.

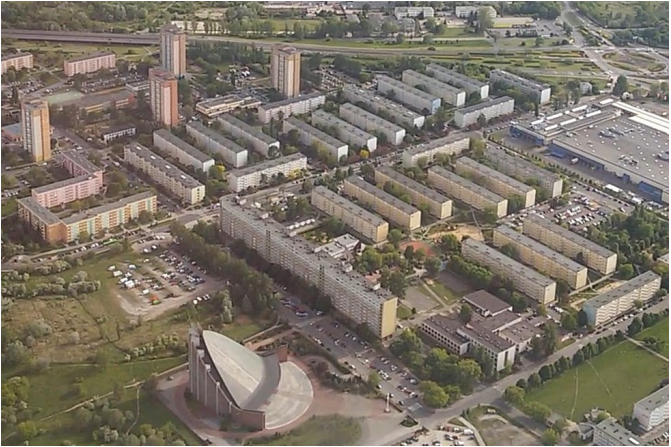

Figure 8. Heroes of the Second World War Subdivision (foreground) and National Army Subdivision (background) in Poznań, Poland

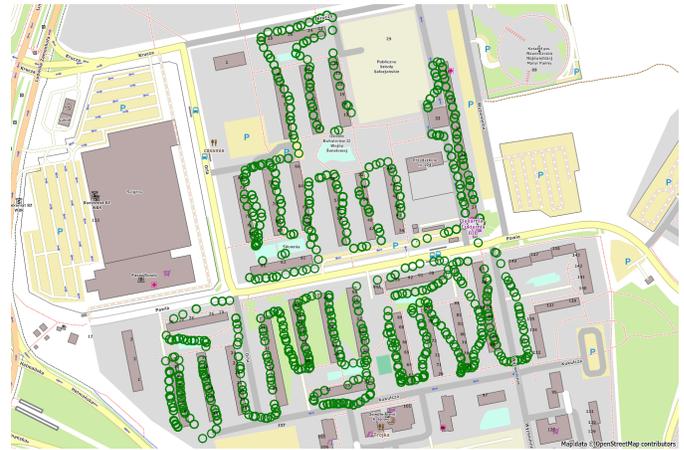

Figure 9. Path of the mobile user in residential hotspot access test

As a representative region of operation, two subdivisions of the Rataje residential area in Poznań, Poland have been selected – Osiedle Armii Krajowej (*English:* National Army Subdivision) and Osiedle Bohaterów II Wojny Światowej (*English:* Heroes of the Second World War Subdivision). Both subdivisions have been built in the 1970s from prefabricated concrete blocks to meet the increasing demand caused by rapid population growth. Most of the constructions are four-story residential blocks (90 to 135 flats each) with a few buildings reaching 10 or 16 storeys, please see Figure 8. According to the 2012 census, the population density in Rataje was 7755,38 person per km$^2$, and hence, the selected subdivisions give a valid overview of the community Wi-Fi concept in one of the largest Polish cities. The results are presented and analysed in the following subsections.

### D. Mobile access evaluation

*1) Experiment methodology:* Mobile access experience has been evaluated with the use of Nexus 5X smartphone (running Android 6.0.1 Marshmallow) that features an IEEE 802.11a/b/g/n/ac (2x2 MIMO) module, as well as, a GPS and GLONASS receiver [16]. During the test the user has been moving on the pavement between the buildings holding the phone in the hand in front of the user. The objective was to record the maximum RSSI (dBm) of every BSSID spotted while traversing the area, together with other parameters of interest, i.e. SSID, AP capabilities, GPS/GLONASS based geographical coordinates, radio frequency, and Wi-Fi channel. Therefore, a free Android application called WiFi Tracker (version 1.2.20) developed by Ian Hawkins [17] has been used to record the data gathered by scanning the radio environment in one second intervals during the whole test. Then the measurements have been exported to a comma-separated values (CSV) file to perform desired analyses.

*2) Results:* In the mobile access test performed on 2 July 2016 around 3 PM, the user has been moving on the pavement between the buildings following the path depicted in Figure 9. Green circles correspond to the geographical coordinates of the location where the maximum RSSI of a BSSID has been recorded. The location discrepancies were caused by the limited GPS/GLONASS visibility between the buildings what has affected the accuracy (a circle on a building or in the middle of a street). The total number of SSIDs and BSSIDs, as well as, the frequencies they were operating on have been presented in Table I, along with the capabilities advertised by every access point.

In the group of 1375 unique SSIDs, 313 distinct `INEA_HotSpot_WiFi` access points have been spotted. In the area of interest, only 8.83% of BSSIDs have been broadcasted at 5 GHz, i.e. 152 out of 1874, with the remaining operating at 2.4 GHz. Most of the BSSIDs (96.48%) advertised that the use of Wi-Fi Protected Access (WPA) and/or Wi-Fi Protected Access II (WPA2) protocol is required to connect. Some of the BSSIDs, i.e. 306 (16.33%), supported Wi-Fi Protected Setup (WPS) security measures. The three most popular 2.4 GHz channels were 1 (34,67%), 11 (28,05%), and 6 (23,40%), as presented in Fig. 10.

The distribution of the maximum RSSI observed for 313 residential `INEA_HotSpot_WiFi` access points has been presented in Figure 11. The maximum signal from 103 APs have been received at the level no weaker than -70 dBm. This value of reference has a practical significance because it is officially used by Apple iOS 8 and later as a threshold for initiating a scan to roam to a different BSSID for the same

Table I
THE GENERAL STATISTICS OF RESIDENTIAL HOTSPOTS MOBILE ACCESS EVALUATION

| | |
|---:|---:|
| `INEA_HotSpot_WiFi` BSSIDs | 313 |
| all SSIDs | 1375 |
| all BSSIDs | 1874 |
| 2.4 GHz BSSIDs | 1722 |
| 5 GHz BSSIDs | 152 |
| WPA/WPA2 protected BSSIDs | 1808 |
| WPS capable BSSIDs | 306 |
| WEP protected BSSIDs | 31 |

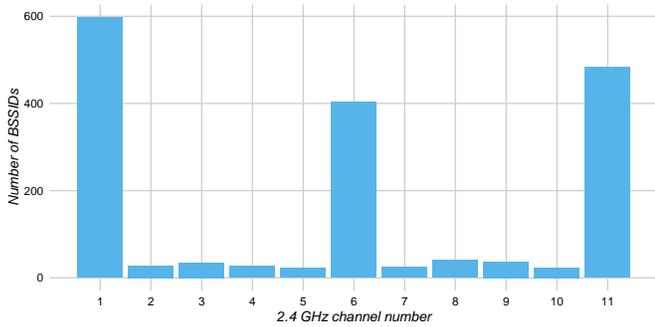

Figure 10. Distribution of 2.4 GHz Wi-Fi channels usage

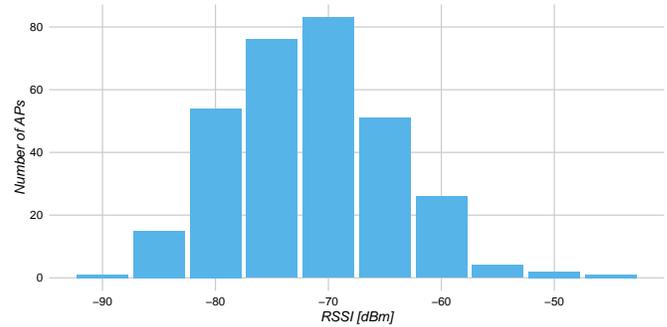

Figure 11. Maximum RSSI distribution of residential `INEA_HotSpot_WiFi` access points

SSID if its signal is at least 8 dB higher (active communication) or 12 dB higher (connection in idle state) [18].

Signal level not lower than -70 dBm can be considered as a value allowing users to enjoy the most popular service types, including voice and video calls. Therefore, because another 81 APs fell into a range between -71 dBm and -75 dBm, empirical tests of a Voice over IP (VoIP) call quality have been performed for those conditions. The calls have been conducted between a user of Nexus 5X and a user of a fixed-line VoIP phone via INEA's commercial VoIP exchange (proxy), with CSipSimple (version 1.02.03 r2457), a popular free VoIP client for Android [19]. The ITU-T G.711A recommendation [20] compliant codec has been used and yielded satisfactory quality with only minor glitches that did not sacrifice the overall experience. As long as the user stayed connected to the BSSID, the call could be continued without interruptions. During the pavement walking test, most of the time the user has been in the radio range of at least one hotspot which could be reached with a required RSSI level. Without the use of dedicated roaming mechanisms, when the user moved out of the radio range and the Wi-Fi connection was lost, the phone started to scan for another BSSID to connect to. Together with the process of 802.1X-based authentication, it can take up to several seconds to re-establish the voice connection and continue the conversation. Therefore, 59% of INEA residential hotspots (184 out of 313) can be used outdoors in the urban environment for VoIP calls, provided that the user stays within the optimum radio range of a single access point (usually a few meters from the nearest building). For the levels lower than -75 dBm, call quality gets worse as signal quality drops, and yet, a user can still benefit from basic web browsing, e-mail and instant messaging (243 community hotspots, i.e. 78% of the total count). The phone starts to keep losing connection to the AP when the RSSI drops to -80 dBm (70 APs have been discovered with RSSI at -80 dBm or worse).

Another real-world test scenario involved a parent engaged in a web browsing session while sitting on a bench by children's playground. In those conditions the quality of experience is rather of best effort type because it heavily depends on the RSSI, and hence, the proximity to the nearest access point. Therefore, according to the experiments, the user will often have to find an optimal spot to seat oneself. Then the user can count on decent quality, what has been additionally proven with a test performed on 21 July 2016 at 10 AM with Speedtest.net application installed on Samsung Galaxy S3 Neo (running Android 4.4.2 KitKat) with the results of 1.94 Mbit/s download, 0.72 Mbit/s upload and 8 ms RTT [21]. Although the test succeeded, in some outdoor conditions, a sufficient link quality will not be possible due to long distance to the nearest AP or difficult radio propagation conditions.

*E. Stationary access evaluation*

*1) Experiment methodology:* The evaluation of stationary access has been performed by measuring the RSSI and the Round Trip Time (RTT) to the IP default gateway, i.e. the aggregation router, as observed by the user wirelessly connected to `INEA_HotSpot_WiFi` access point located in different apartment in the same building. A laptop running Windows 7 equipped with an IEEE 802.11n card has been chosen as a test platform. The sampling has been conducted in the period of one week during the average busy-hour of the Internet access services in INEA in the summer, i.e. between 9:30 PM and 10:30 PM. This part of the day has been selected based on the assumption that it will provide the most challenging RF conditions (noise, interference, medium congestion etc.) for the residential Wi-Fi usage. RSSI has been recorded every 2 seconds with Homedale (version 1.61) [22], a free Wi-Fi scanner. The ICMP-based RTT has been measured every 1 second with hrPing (version 5.06) [23], a feature-rich freeware ping utility.

*2) Results:* The mean results of busy-hour measurements are presented in Figure 12, where the solid line corresponds to mean RSSI and the dashed line corresponds to mean RTT. The mean RSSI for every busy-hour oscillated between -63 and -68 dBm with the confidence intervals (error bars) being not visible since the widest was only 0.146 dBm. On the other hand, the confidence intervals for mean RTT are distinguishable because sample series are significantly varied.

As validated with the tests, the key to drawing valid conclusions from the analysis of the RTT samples in suboptimal radio conditions is focusing not (only) on the mean values, even when thousands of samples are available, but (also) on the wider scope of statistical parameters, as analysed in Table II. As it turns out, a pretty steady mean RTT that varies from 20.508 to 28.573 ms can be accompanied by a significant range of standard deviation, i.e. between 25.190 and 33.643 ms. Because RTT takes only positive values, that relation suggests that there is not only a wide variation, but also an asymmetry in the distribution. As it turns out, in the studied case there are fewer samples of high values, with some as high as 992.670 ms, while the most frequent ones were as low as only 5.422 ms. As cumulatively depicted in Figure 13, the distribution of RTT throughout 7 days examination is visibly long-tailed with 69.55% of samples not higher than 20 ms, 74.92% not higher than 30 ms and 94.55% not exceeding 80 ms.

Table II
ROUND TRIP TIME (RTT) RELATED STATISTICS OF RESIDENTIAL INEA_HotSpot_WiFi ACCESS POINT STATIONARY TEST (IN MILISECONDS)

| day | 1 | 2 | 3 | 4 | 5 | 6 | 7 |
|---|---|---|---|---|---|---|---|
| mean RTT | 22.381 | 22.606 | 26.408 | 28.573 | 20.508 | 24.469 | 22.722 |
| confidence interval | 1.121 | 1.123 | 1.448 | 1.425 | 1.116 | 1.512 | 1.082 |
| standard deviation | 26.092 | 26.132 | 33.643 | 33.159 | 25.960 | 35.154 | 25.190 |
| minimum | 5.422 | 5.509 | 5.535 | 5.549 | 5.618 | 5.741 | 5.711 |
| maximum | 310.987 | 318.919 | 525.134 | 731.002 | 614.059 | 992.670 | 197.314 |
| mean one way jitter | 10.492 | 10.450 | 13.133 | 13.702 | 9.211 | 11.830 | 10.339 |
| second quartile (median) | 3.043 | 2.983 | 5.340 | 7.081 | 1.851 | 3.232 | 3.007 |
| third quartile | 19.127 | 18.215 | 22.916 | 23.075 | 14.875 | 20.445 | 19.233 |
| maximum | 144.459 | 154.386 | 253.140 | 346.468 | 303.031 | 467.485 | 94.988 |

Since a variation in the RTT is visible, it is of key importance to investigate how exactly, from the user's perspective, the delay changes in time, so in other words, to scrutinize the jitter. Hence, the focus should be put on one way jitter, as one of the key parameters used by network designers and service providers to assess the quality of real time services. One way jitter presented in Table II has been calculated as the half of the mean absolute difference between every two consecutive RTT samples, yielding results between 9.211 ms and 13.702 ms. In most cases, these values satisfy the rule of a thumb that recommends the end-to-end jitter for VoIP communication not to exceed 30 ms [24]. Even if we double the results for the sake of the other party operating in similar last-mile conditions and leave a few ms safety margin for the jitter in the backbone networks, the call quality should still meet the expectations, especially when enhanced by voice buffering algorithms. Although, seeing that the averages may not give the full picture, the second and third quartiles should be of interest. The second quartile (the median) shows that 50% of the differences between two samples are lower than single milliseconds. Nevertheless, the third quartile indicates that the highest 25% of differences have exceeded 23.075 ms with the highest one reaching 467.485 ms, what might have led to a glitch if a voice call had been taking place.

Although the stationary access evaluation is based on a particular case of synthetic tests, it gives a solid overview of what can be expected in similar conditions. Aside from the less delay and jitter demanding 2 Mbit/s web browsing, a seamless voice call might not be possible in every usage scenario related to INEA_HotSpot_WiFi service, but certainly there is a degree of freedom a community member can enjoy and use the service also outside the apartment of a fellow member. Moreover, the community Wi-Fi can be considered a useful Wi-Fi offload solution, enabling users to use the local radio communication, instead of transferring the data through 3G or LTE networks. Still, the further from the nearest community Wi-Fi access point, the higher the likelihood of the packet drops and delay variations.

## IV. CONCLUSIONS

Little is known about performance of commercially available hotspot networks. Therefore, this article shares practical observations of a versatile service provider and presents the analyses performed in mobile and residential INEA Wi-Fi hotspot network in comparison with INEA stationary 5 GHz Wi-Fi access network. The paper reports results of experiments performed over the period of one month. Some trends are clearly visible, such as, the peaks in user number in the rush hours (mobile hotspots) and in the afternoon (residential hotspots), or the general characteristics of noise floor distribution, regardless of the type or location of an access point. Moreover, a survey of 2.4 GHz channels usage has been presented with the statistics of Wi-Fi security measures deployed in urban environment. Community Wi-Fi user experience evaluation has been conducted and discussed, proving that the service fulfils the requirements of VoIP transmission and web browsing. The synergy achieved by the combination of two types of hotspot network provides coverage not only in the city centre but also in more remote urban areas reached by INEA cable network.

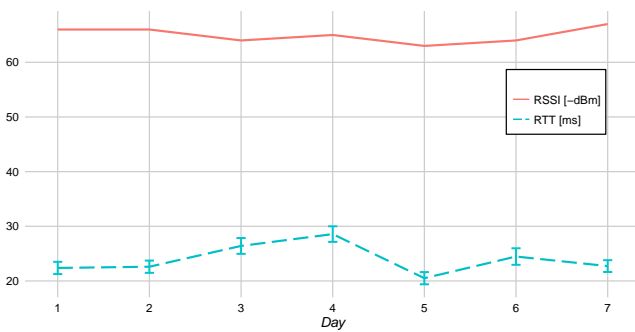

Figure 12. Mean values of RSSI and RTT to the gateway throughout 7 days of stationary test of a residential INEA_HotSpot_WiFi access point

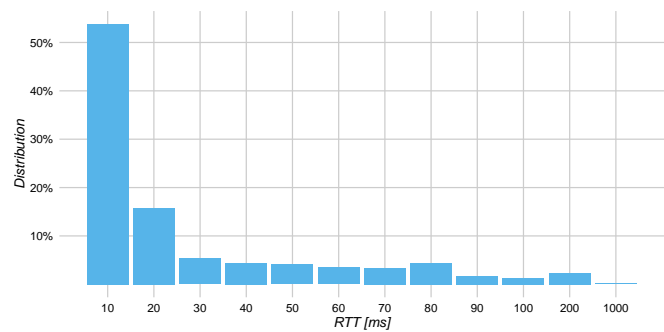

Figure 13. Distribution of RTT measured to the gateway in stationary test of a residential INEA_HotSpot_WiFi access point